%% file: ms.tex
\documentclass[12pt]{aastex}
\usepackage{graphicx}
\usepackage{natbib}
\usepackage{longtable}
\usepackage{amsmath}
\usepackage{subfigure}
\usepackage{placeins}
\begin{document}
\title{The Gould's Belt Very Large Array Survey. III. \\ The Orion region}
\author{Marina Kounkel\altaffilmark{1}, Lee Hartmann\altaffilmark{1}, Laurent Loinard\altaffilmark{2,3}, Amy J. Mioduszewski\altaffilmark{4}, Sergio A. Dzib\altaffilmark{3}, Gisela N. Ortiz-Le\'{o}n\altaffilmark{2}, Luis F. Rodr\'{i}guez\altaffilmark{2}, Gerardo Pech\altaffilmark{2}, Juana L. Rivera\altaffilmark{2}, Rosa M. Torres\altaffilmark{5}, Andrew F. Boden\altaffilmark{6}, Neal J. II Evans\altaffilmark{7}, Cesar Brice\~{n}o\altaffilmark{8}, John Tobin\altaffilmark{9}}

\altaffiltext{1}{Department of Astronomy, University of Michigan, 500 Church Street, Ann Arbor, MI
48109, USA}
\altaffiltext{2}{Centro de Radiostronom\'{i}a y Astrof\'{i}sica, Universidad Nacional Aut\'{o}noma de Mexico, Morelia 58089, Mexico}
\altaffiltext{3}{Max Planck Institut f\"{u}r Radioastronomie, Auf dem H\"{u}gel 69, 53121 Bonn, Germany}
\altaffiltext{4}{National Radio Astronomy Observatory, 1003 Lopezville Road, Socorro, NM 87801, USA}
\altaffiltext{5}{Instituto de Astronom\'{\i}a y Meteorolog\'{i}a, Universidad de Guadalajara, Avenida Vallarta No. 2602, Col. Arcos Vallarta, CP 44130, Guadalajara, Jalisco, M\'{e}xico}
\altaffiltext{6}{Division of Physics, Math and Astronomy, California Institute of Technology, 1200 E California Blvd., Pasadena, CA 91125, USA}
\altaffiltext{7}{Department of Astronomy, The University of Texas at Austin, 1 University Station, C1400, Austin, TX 78712, USA}
\altaffiltext{8}{Cerro Tololo Interamerican Observatory, Casilla 603, La Serena, Chile}
\altaffiltext{9}{National Radio Astronomy Observatory, Charlottesville, VA 22903, USA}

\clearpage

\begin{abstract}
We present results from a high-sensitivity ($~$60 $\mu$Jy), large-scale (2.26 square degree) survey obtained with the Karl G. Jansky Very Large Array as part of the Gould's Belt Survey program. We detected 374 and 354 sources at 4.5 and 7.5 GHz, respectively. Of these, 148 are associated with previously known Young Stellar Objects (YSOs). Another 86 sources previously unclassified at either optical or infrared wavelengths exhibit radio properties that are consistent with those of young stars. The overall properties of our sources at radio wavelengths such as their variability and radio to X-ray luminosity relation are consistent with previous results from the Gould's Belt Survey. Our detections provide target lists for followup VLBA radio observations to determine their distances as YSOs are located in regions of high nebulosity and extinction, making it difficult to measure optical parallaxes.
\end{abstract}

\clearpage

\section{Introduction}

The Orion Molecular Clouds are one of the nearest active star-forming regions, containing several thousand pre-main sequence stars as well as a substantial number of massive stars including the O7 binary $\theta^1$ C Ori in the center of the Orion Nebula Cluster (ONC). The stellar populations in the two main Orion A and B clouds can be divided into distinct clusters such as the ONC, OMC 2/3, NGC 2023, 2024, 2068 and 2071, and Lynds 1622, along with the more distributed population in the L1641 region \citep{ori, onc, ngc2023, ngc2068, l1622, l1641, tina}. The $\lambda$ and $\sigma$ Ori clusters, while not part of the Orion A and B clouds, are also considered to be part of the Orion OB1 association \citep{laori, sigori}.

In order to fully understand the history of star formation in Orion, it is important to determine accurate distances as well as kinematics. The spatial positions, three-dimensional velocities, and ages of the stars are essential to testing theories of star cluster formation. However, accurate distance measurements to these stars are needed, as uncertainties of 20\% or more in the distance can translate into errors of 40\% in total brightness or luminosity, resulting in a 70\% error in ages for low-mass stars \citep{hartmann2001}. 

The distance to the Orion Complex has been a subject of much debate over the years, with accepted distances ranging from 380 to 520 parsecs \citep{sfrbook}. It is only recently that a firmer estimate of 414$\pm$7 pc, obtained by using high-resolution radio interferometry of four stars in the central regions of the ONC (the Trapezium), has come to be accepted \citep{menten2007}. However, these four stars may not be representative of the main cluster, much less of the entire Orion A/B complex that spans a range projected on the sky of $\sim 100$ pc. Furthermore, it is known that stars from a somewhat older group, the Orion Ia association, lie in front of the main Orion A molecular cloud by as much as 50 -- 100 pc \citep{ori, alves2012}, confusing attempts to identify the true young stellar population.

The most direct method to determine the distance of a star is through observing its parallax. However, it becomes increasingly difficult to determine at large distances and toward regions with high nebulosity. For an object located 500 pc away the parallax is only 2 milli-arcseconds; to determine the distance to an accuracy of 2\% a measurement error of only 40 micro-arcseconds is required. Such accuracy cannot be obtained with current ground-based optical techniques, or from the Hubble Space Telescope, but can be achieved using Very Long Baseline Interferometry at radio wavelengths. The Very Long Baseline Array (VLBA) is one of the largest interferometers with 10 different sub-arrays located across the United States and a maximum effective baseline of 8,611 km. It can produce such accuracy and can complement \textit{Gaia} in regions of star formation, where extinction and nebulosity pose problems for optical surveys.

Since VLBA has a very high spatial resolution, sources need to have high surface brightness in order to be detected. Therefore, good candidates are those that are suspected to emit non-thermally at radio frequencies. Not all young stars are non-thermal emitters, so it is necessary to perform a survey at lower resolution to identify candidates and rule out background quasars. In particular, objects classified as Class III stars based on their infrared SED tend to exhibit gyrosynchrotron non-thermal emission that is related to magnetic activity of the star. This emission is frequently associated with a high degree of circular polarization and variability \citep{dulk1985}.

The \textit{Gould's Belt Very Large Array Survey} is an ongoing large-scale effort to map all the neighboring star-forming regions in radio frequencies and to identify likely non-thermal radio emitters. These stars will be later used to determine accurate distances and three-dimensional structure and kinematics across all regions using VLBA \citep{Dzib2013, Gisela}. In this paper, which is the third in the series, we focus on Very Large Array observations of the Orion Molecular Cloud Complex. In Section \ref{sec:obs} we describe the observation details and in Section \ref{sec:result} we present an overview of the detected objects. We analyze the source properties in Sections \ref{sec:dis} and \ref{sec:sour} and summarize our conclusions in Section \ref{sec:concl}.

\section{Observations} \label{sec:obs}
\begin{figure}[c]
  \centering
    \includegraphics{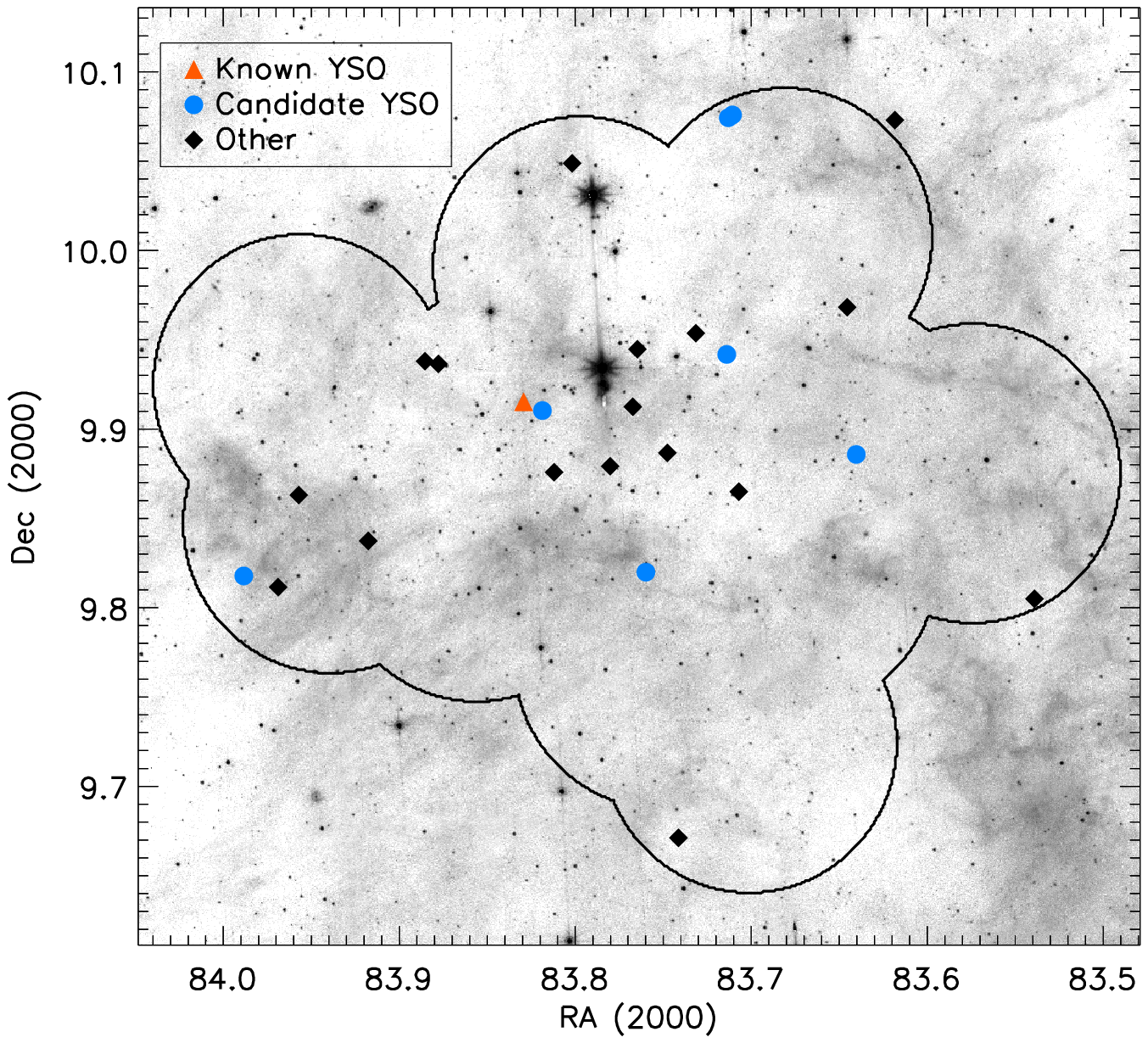}
      \caption{The $\lambda$ Ori field. The 8.0 $\mu$m Spitzer map is taken from \cite{Barrado2007}. The outline shows the radio coverage of the field with VLA at 4.5 GHz with FWHM diameter of the primary beam of 10'; symbols show the positions of the detected objects. Red triangles represent objects that have been identified as YSOs in previous surveys; blue circles are candidate YSOs based on their radio properties; black diamonds are all the remaining objects. See Section \ref{sec:result} for a description of each category. \label{fig:la}}
\end{figure}

\begin{figure}[c]
  \centering
    \includegraphics{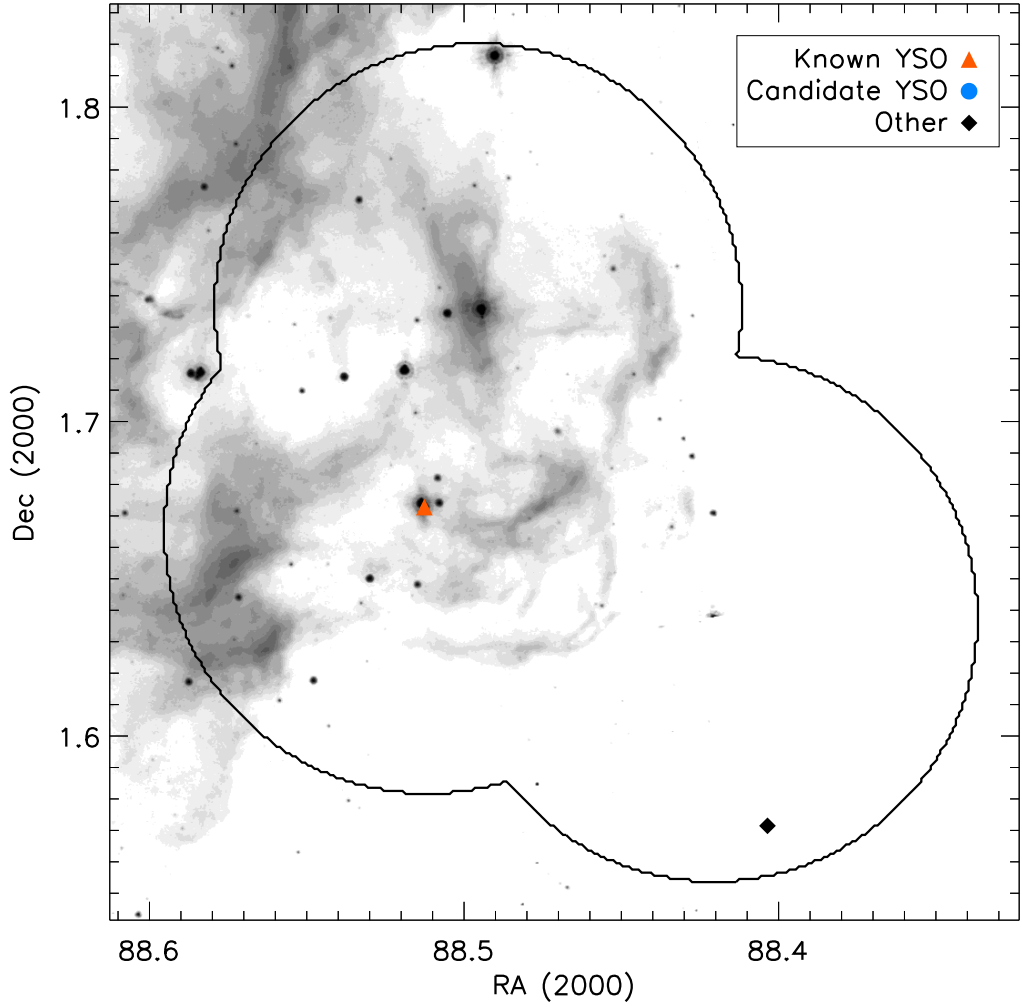}
      \caption{Same as Figure \ref{fig:la}, but showing the Lynds 1622 region. The 8.0 $\mu$m Spitzer map is taken from \cite{Megeath2012}. \label{fig:l1622}}
\end{figure}

\begin{figure}[c]
  \centering
    \includegraphics{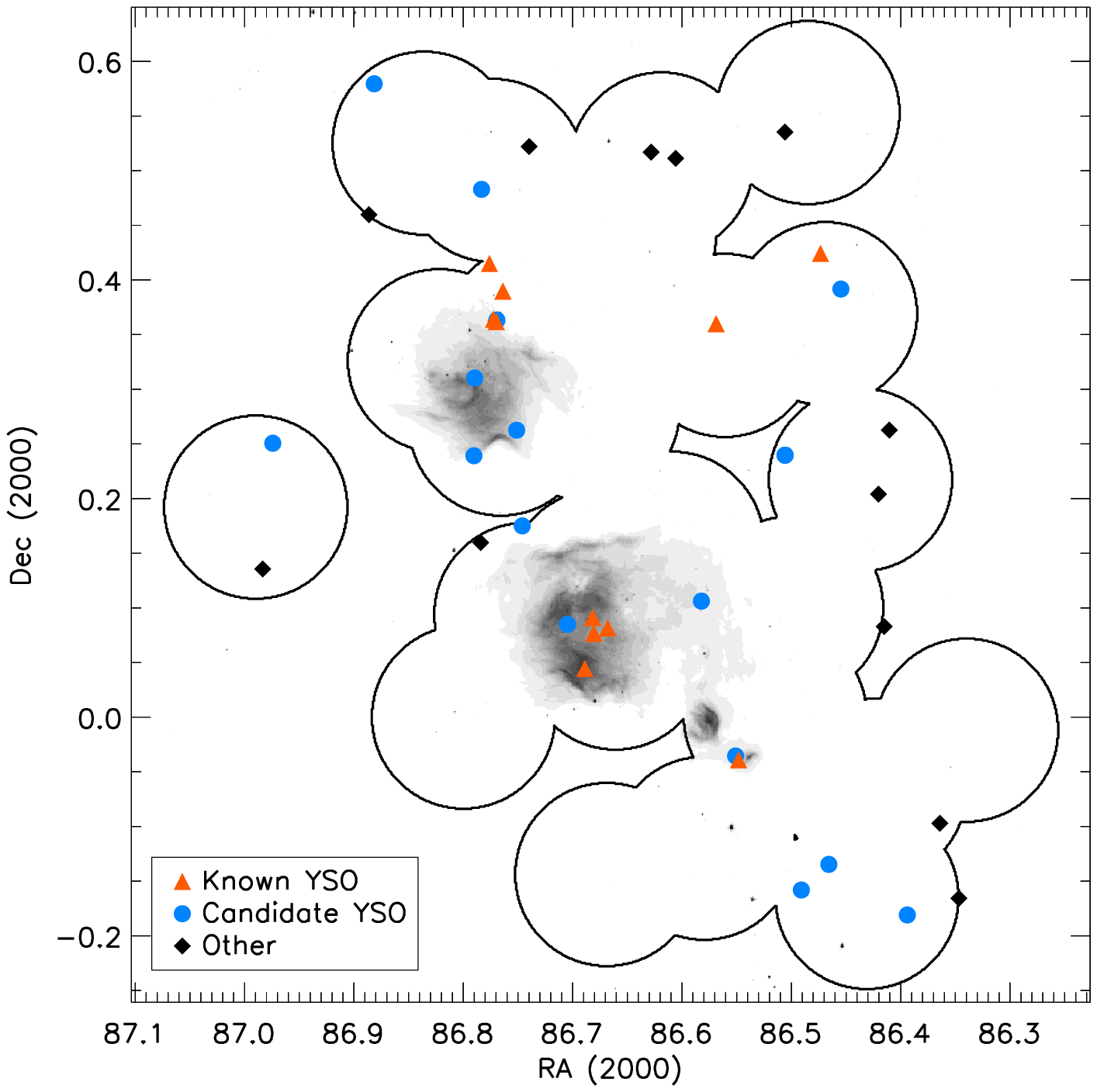}
      \caption{Same as Figure \ref{fig:l1622}, but showing the NGC 2068 and NGC 2071 regions. \label{fig:n2068}}
\end{figure}

\begin{figure}[c]
  \centering
    \includegraphics[height=0.9\textheight]{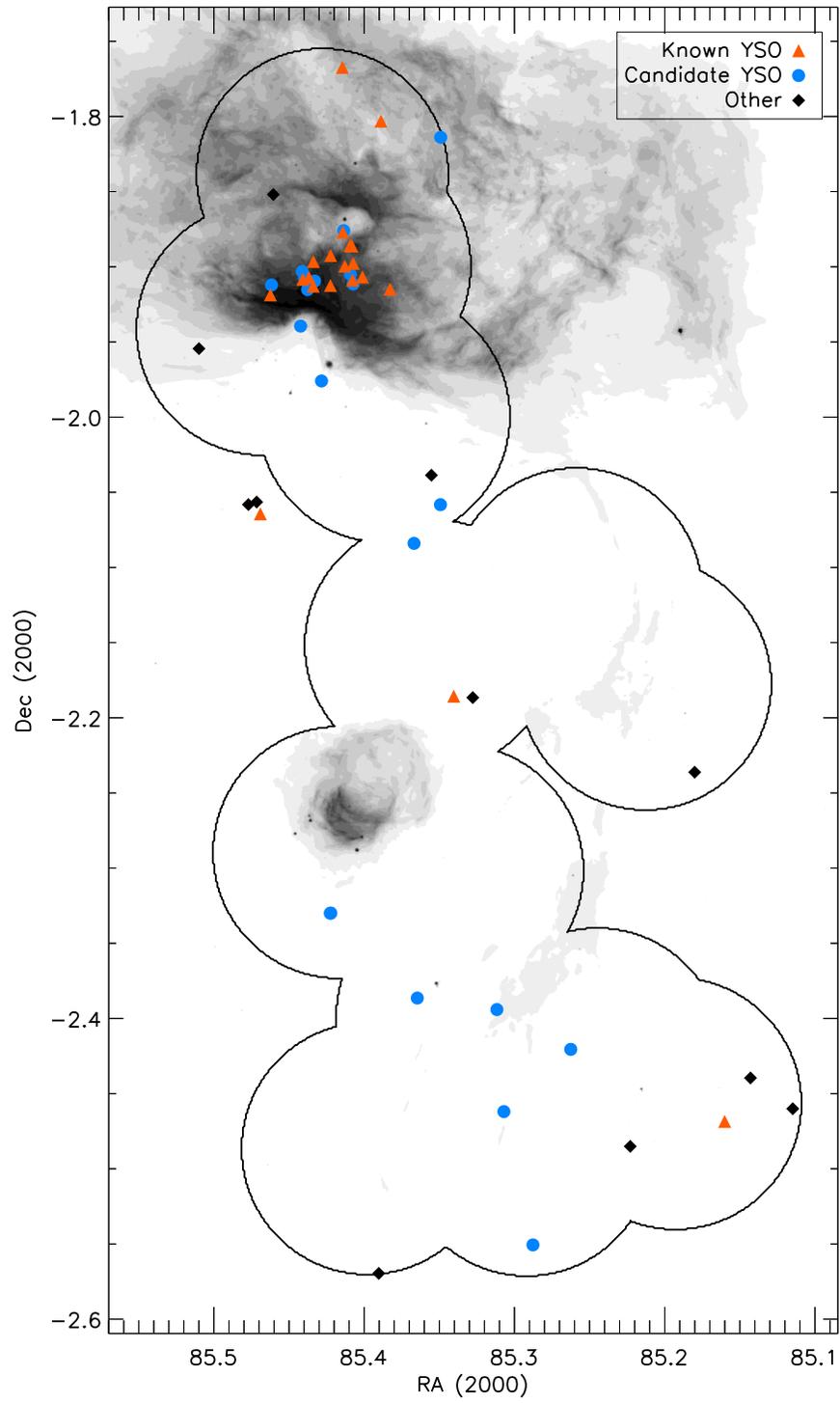}
      \caption{Same as Figure \ref{fig:l1622}, but showing the NGC 2023 and NGC 2024 regions. \label{fig:n2024}}
\end{figure}

\begin{figure}[c]
  \centering
    \includegraphics{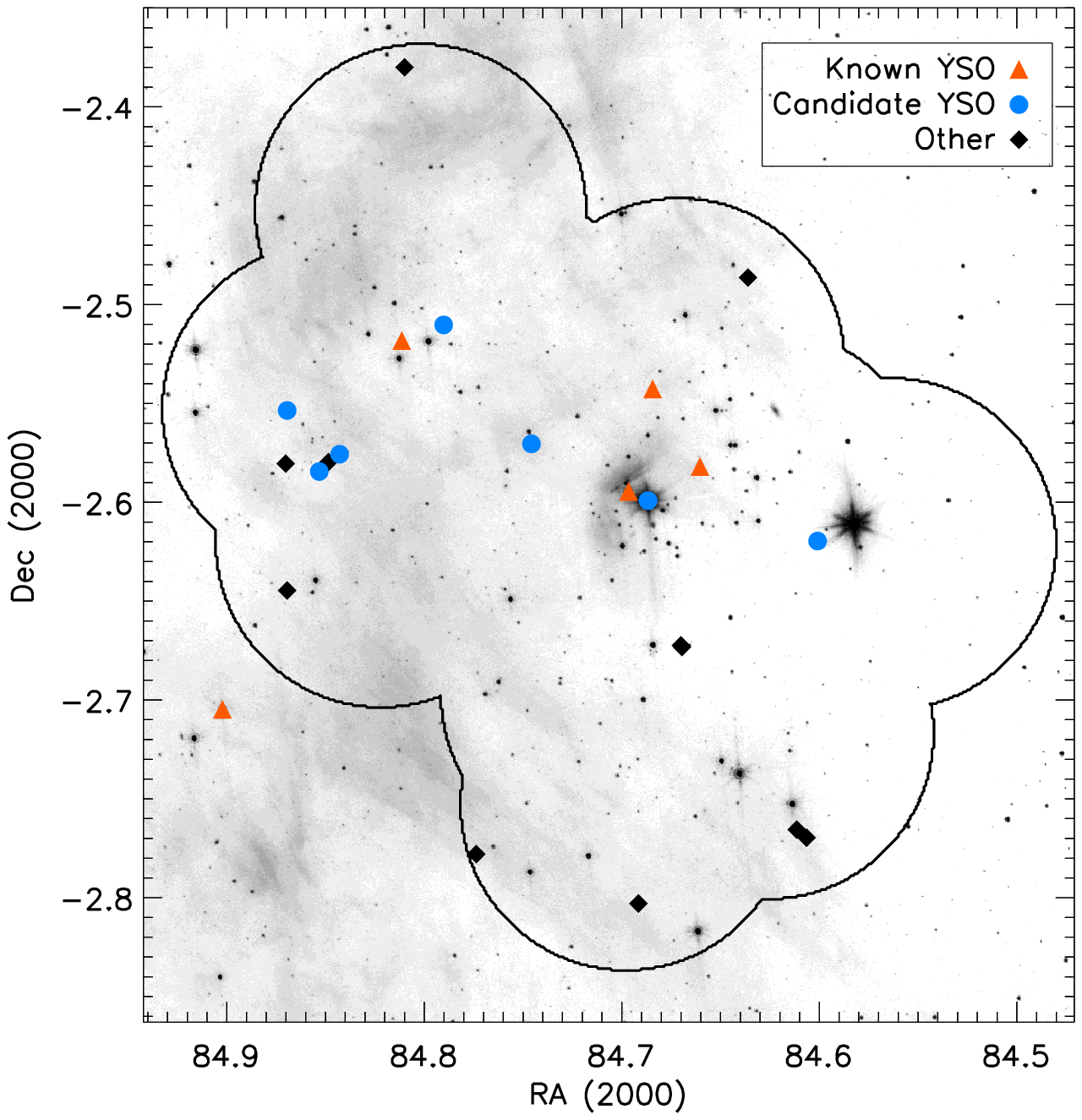}
      \caption{Same as Figure \ref{fig:la}, but showing the $\sigma$ Ori field. 8.0 $\mu$m Spitzer map is taken from \cite{Hernandez2007}. \label{fig:si}}
\end{figure}

\begin{figure}[c]
  \centering
    \includegraphics[height=0.9\textheight]{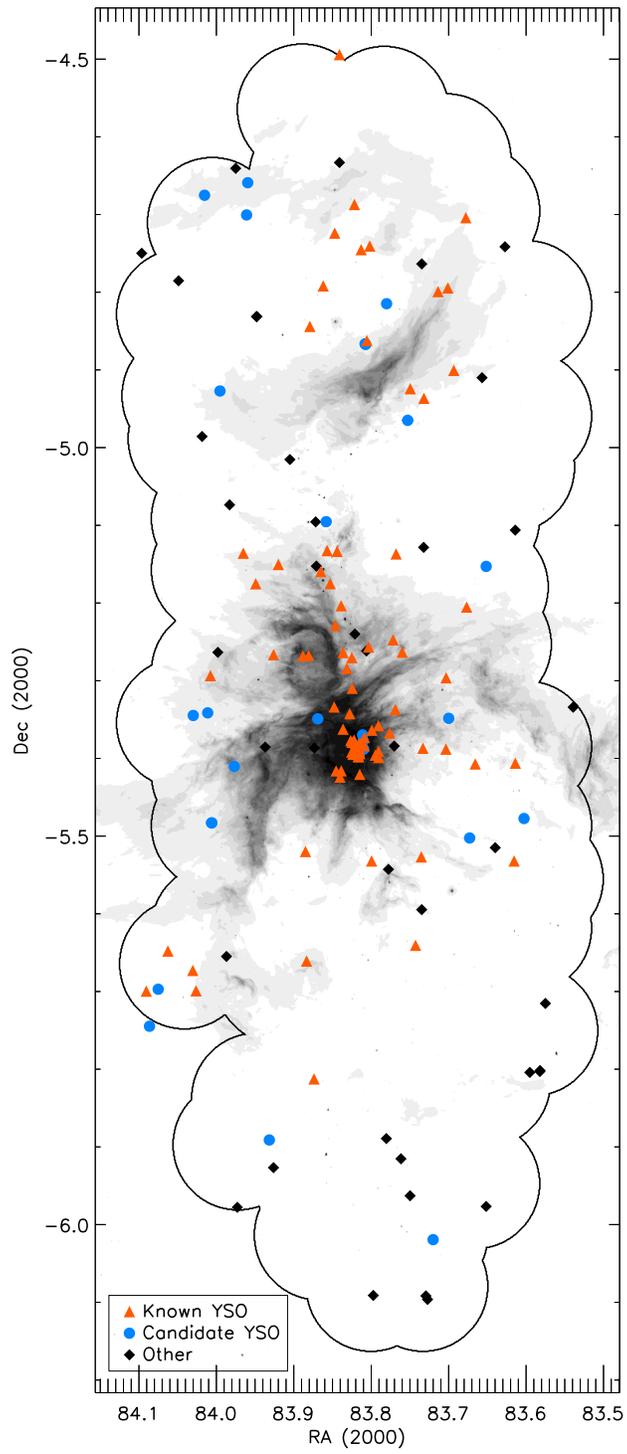}
      \caption{Same as Figure \ref{fig:l1622}, but showing the Orion Nebula Cluster region. \label{fig:onc}}
\end{figure}

\begin{figure}[c]
  \centering
    \includegraphics{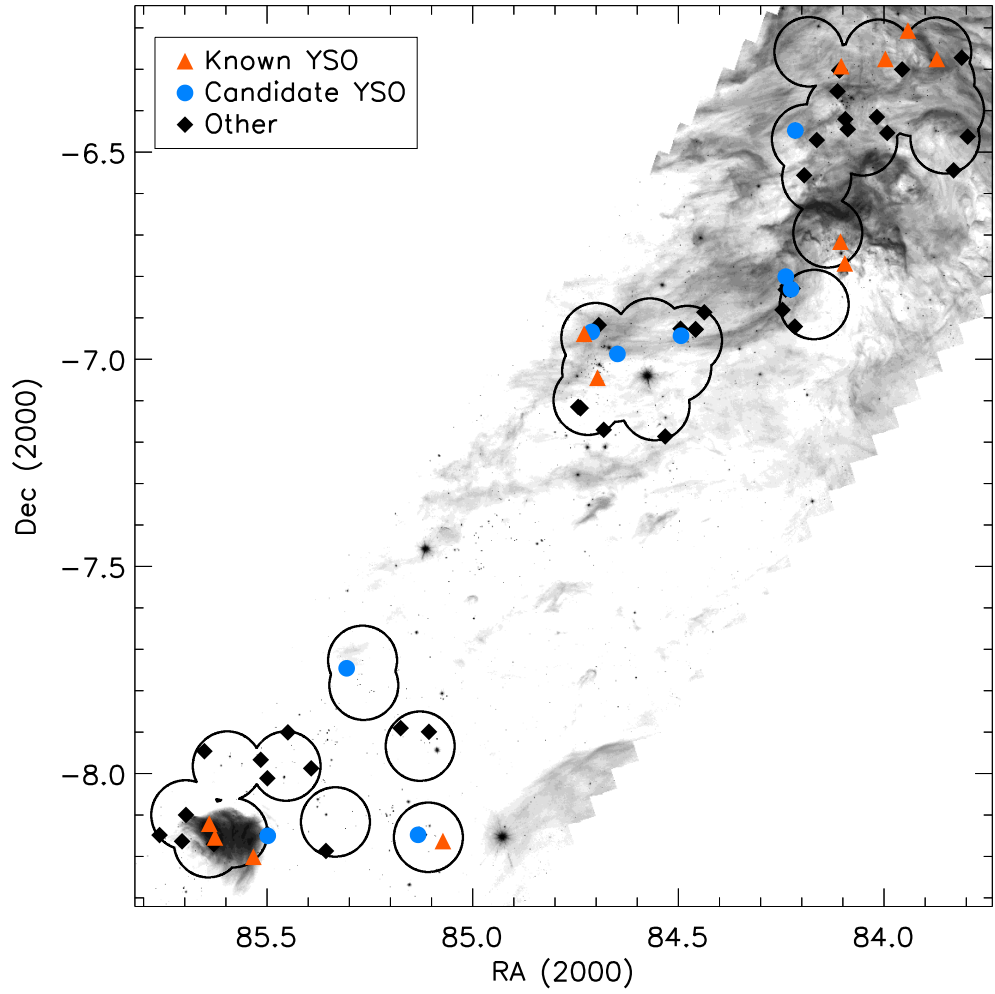}
      \caption{Same as Figure \ref{fig:l1622}, but showing the Lynds 1641 region. \label{fig:l1641}}
\end{figure}

Fields in the Orion A and B molecular clouds were observed with the Karl G. Jansky Very Large Array (VLA) in its A configuration. The 210 individual fields have been split into seven maps, with 30 fields being observed per map, as follows: 12 in $\lambda$ Ori, 3 in L1622, 27 are shared between NGC 2068 and NGC 2071, 14 are shared between NGC 2023 and NGC 2024, 11 in $\sigma$ Ori, 109 in the Orion Nebula Cluster (ONC), 16 in L1641-N, 8 in L1641-C, and 10 in L1641-S (see Figures \ref{fig:la}-\ref{fig:l1641}). All the maps were imaged closely in time, and a total of three epochs separated by approximately a full month were acquired in summer 2011 (Table \ref{tab:Dates}). Two 1 GHz frequency bands were observed simultaneously, at 4.5 and 7.5 GHz. Fields were positioned in such a way as to provide a uniform coverage over the extended area surrounding the known positions around young stars. Assuming a FWHM diameter of the primary beam of 10' and 6' at 4.5 and 7.5 GHz respectively, the total area covered by our observations is 2.26 and 1.35 square degrees, respectively.

3C 138 was used as the flux calibrator for all the fields. Three phase calibrators were observed: J0532+0732 for $\lambda$ Ori; J0552+0313 for L1622, NGC 2068; J0541-0541 for NGC 2023, 2024, $\sigma$ Ori, ONC, and all of the L1641 fields. The observational setup was the same as described in \cite{Dzib2013}. Data were reduced and analyzed using Astronomical Image Processing System (AIPS). Images of individual fields were constructed and corrected for the primary beam response in a standard fashion separately for all three epochs at both 4.5 and 7.5 GHz. 

We achieved a nearly uniform rms noise of 60 $\mu$Jy beam$^{-1}$ at both frequencies in all the regions. The only exception to this is in the Trapezium region due to nebular emission; there the noise was 200 $\mu$Jy beam$^{-1}$ after excluding baselines smaller than 150 k$\lambda$ during imaging to remove extended emission.

\section{Results}\label{sec:result}

\begin{figure}
        \centering
			\subfigure{\includegraphics[width=0.45\textwidth]{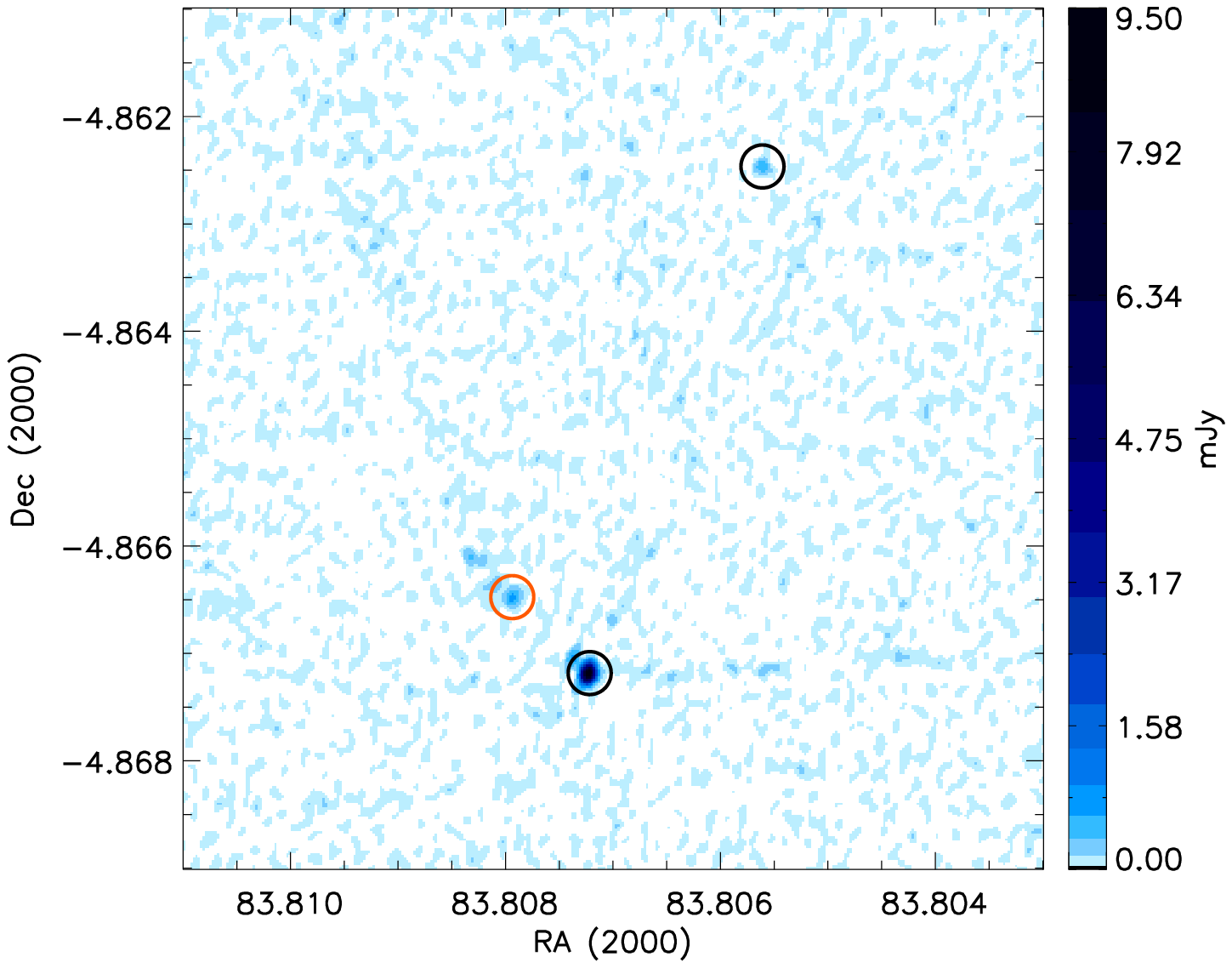}}
    			\subfigure{\includegraphics[width=0.45\textwidth]{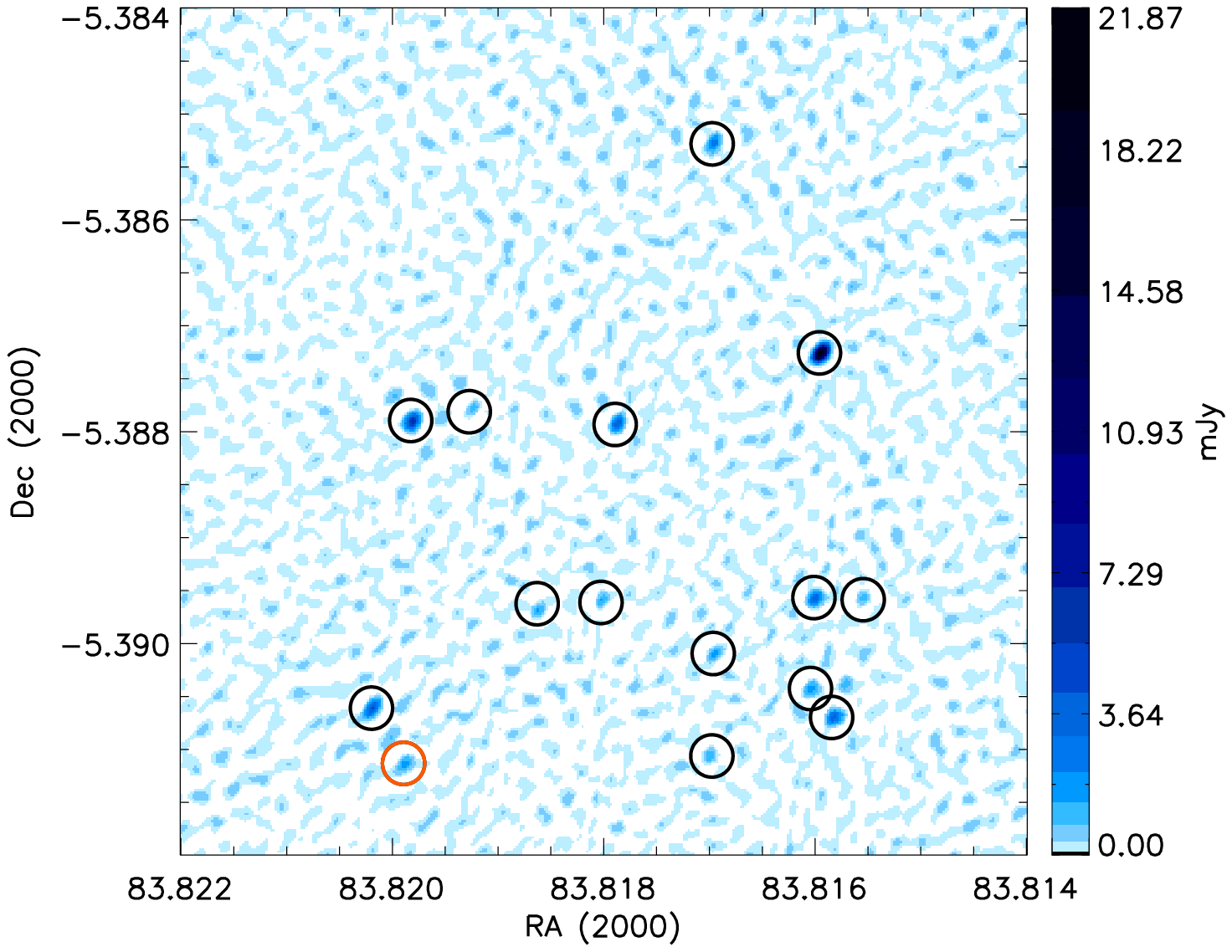}}    
        \caption{Example of a single epoch cleaned 4.5 GHz VLA map of a region with typical noise level (left) and a high noise region such as the Trapezium (right). Identified sources are circled. Black circles show objects that have been previously identified in the various multi-wavelength surveys, orange are the new detections. \label{fig:imagr}}
\end{figure}

Sources were identified through a visual inspection of the individual fields at 4.5 GHz during the cleaning and imaging process since an automated source identification was deemed to be not sufficiently advanced and produced results that were too unreliable. An example of produced images is shown in the Figure \ref{fig:imagr}. In particularly clustered regions such as Trapezium and NGC 2024, in addition to standard imaging, data from all three epochs were combined into a single image for source identification purposes only to improve statistical significance of each detection. Rms noise in the vicinity of an object was extracted using IMSTAT over a region of size 10,000 -- 100,000 pixels.

We have detected a combined total of 374 sources among three epochs for all the regions (Table \ref{tab:all}). Since they were taken as part of \textit{Gould's Belt Very Large Array Survey}, we assign them a name of GBS-VLA J\textit{hhmmss.ss-ddmmss.s}, where J\textit{hhmmss.ss-ddmmss.s} is the J2000.0 coordinate of each source.

Fluxes at 4.5 and 7.5 GHz were measured by performing a two-dimensional Gaussian fitting for each object in all three epochs using JMFIT. We consider two sources of uncertainty in flux --- statistical noise in the images and a systematic uncertainty of 5\% from possible errors in the absolute flux calibration. We also present the calculated spectral index $\alpha$ defined such that flux dependency on frequency is $S_\nu \propto \nu^{\alpha}$.

All sources but one had fluxes greater than five times the rms noise in at least one epoch. The remaining source, GBS-VLA J053518.67$-$052033.1, was detected at two epochs with maximum detection probability of 4.9$\sigma$ in a single epoch data. It is found in the Trapezium region, and has known counterparts in other wavelength regimes.

\begin{figure}[c]
  \centering
    \includegraphics{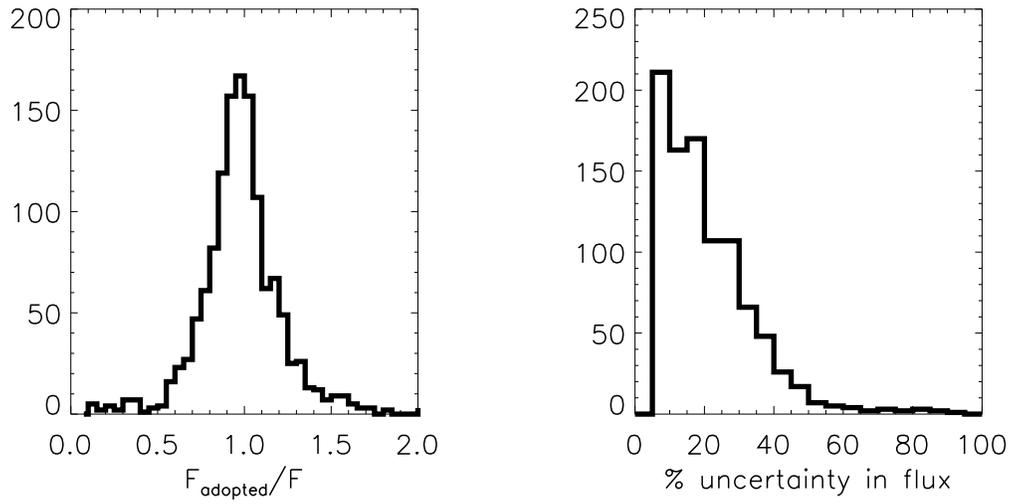}
      \caption{Left: Histogram of fraction of adopted flux (detection closest to the beam center) to flux of all the other detections of the same source at 4.5 GHz. Right: The percentage of measured and systematic 5\% uncertainties added in quadrature relative to the measured flux. Both plots include data from all three epochs. Sources with a high degree of uncertainty in the flux have been identified at positions corresponding to a more confident detection in at least one of the epochs.\label{fig:unc}}
\end{figure}

Since our fields have been positioned in a way to provide uniform sensitivity, there was significant overlap between them. Therefore, for many of our sources we have several detections at different positions on the beam within the same epoch. Whenever this was the case, we selected a detection in which a source would be closest to the beam center to provide the adopted flux for the epoch. For each source, we present only the epoch with the largest flux at each frequency. Based on the overlap, we determine that our coordinates are generally accurate to $<$0.2$"$. Similarly, total flux uncertainty is on the order of 20\%, which is recovered by the combined measured and systematic uncertainties in the data (Figure \ref{fig:unc}).

We calculated the variability as the difference between highest and lowest measured flux, normalized by the maximum flux. Uncertainties in the variability were calculated by adding statistical and systematic errors in quadrature for both epochs and combined using error propagation. These percentages are quoted for 4.5 and 7.5 GHz in columns 3 and 5 of Table \ref{tab:all}. For uncertainties in variability, we also consider uncertainty from the pointing error of VLA primary beam as described by \cite{dzib2014}. While the coordinate grid itself is largely unaffected, the location of the phase center of the primary beam itself is usually uncertain to 10 -- 20$''$ \citep{rupen1997}, which results in an inconsistent response that becomes particularly important in the outer edges of the field. For an object located 3$'$ from the center, this could lead to 8\% uncertainty in variability at 4.5 GHz and 23\% at 7.5 GHz.

We identified true variable sources as those exhibiting a change in flux greater than 50\% at either frequency. Additionally, we also considered sources to be variable if they were not detected at one or more epochs and were located in the inner half of the beam. Sources for which circular polarization was confidently detected are listed in Table \ref{tab:pol}.

We cross-referenced our catalog of sources with previous major radio, infrared, optical and X-ray surveys of the regions published in the literature (Table \ref{tab:known}). We have generally considered sources in these surveys to be counterparts if they had positional coincidences less than 1$"$, but have allowed for larger offsets if the combined uncertainty between the databases was large.

\begin{figure}[c]
  \centering
    \includegraphics{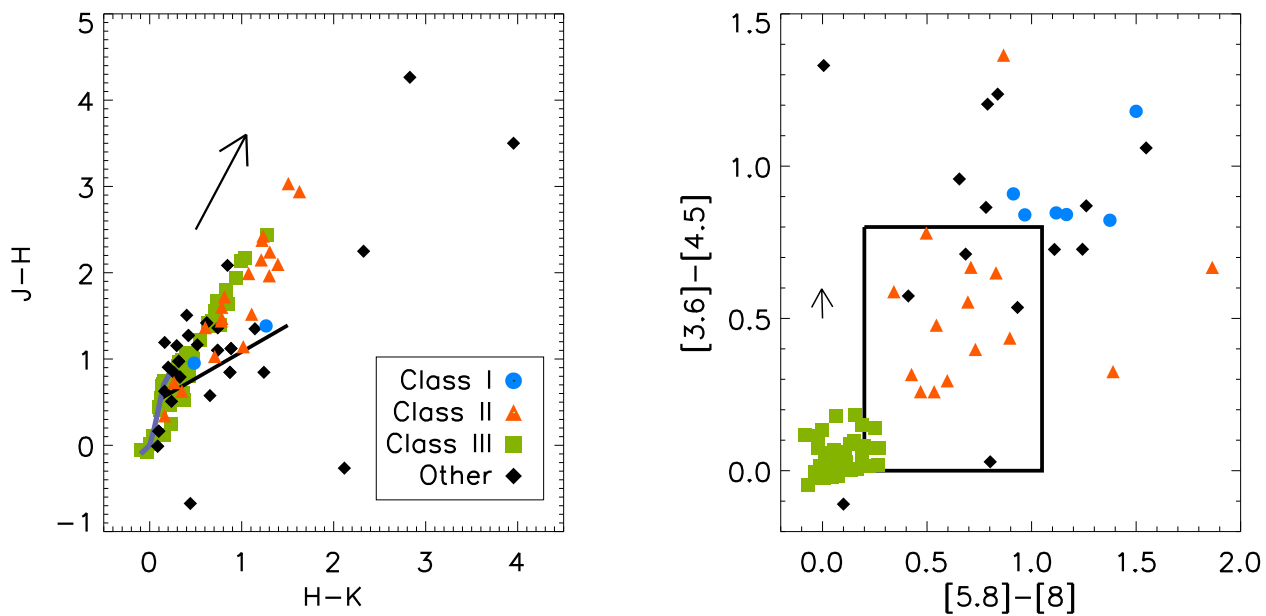}
      \caption{Color-color diagrams for sources with infrared counterparts from Table \ref{tab:known}, identified according to their evolutionary classification, when known. Arrows show the reddening vector of 1 $A_K$, from \cite{Megeath2012}. On the left plot, a black line shows the location of the CTTS locus as identified by \cite{meyer1997}, which corresponds to the intrinsic de-reddened colors of the young stars with disks. A purple line shows typical colors for pure photospheres \citep{bessell1989}. On the right plot, the rectangle shows the approximate colors of Class II stars as identified by \cite{allen2004}\label{fig:color}}.
\end{figure}

Of 374 detected sources, 261 have been previously found at another wavelength region, while 113 are new detections. 146 sources have been detected in X-rays, 94 at optical wavelengths, 218 at infrared, and 63 in previous radio surveys. For sources with infrared counterparts we display infrared color-color diagrams in Figure \ref{fig:color}. Of the previously identified sources, one is extragalactic, while the other 148 as young stellar objects (YSOs). Of the YSOs, 106 have been placed on the standard class system based on the IRAC color-color classification of \cite{allen2004}. There are 11 Class 0/I, 26 Class II, and 70 Class III type stars (Table \ref{tab:known}).

A total of 225 sources are either new detections or, to our knowledge, have not been previously classified in the literature. Of these remaining objects, we have identified 86 as exhibiting variability or high levels of circular polarization (Table \ref{tab:new}). While we cannot exclude the possibility that any of them are extragalactic in nature, quasars are not expected to vary as strongly on time scales of few weeks to few months \citep{hovatta2008}, and exhibit very weak circular polarization \citep{saikia1988}, so these sources are likely YSO candidates. Using the same criteria of variability and circular polarization would identify only 107 of the 148 previously-known YSOs; thus we cannot tell which of the remaining 139 unidentified sources are YSOs or extragalactic objects. Further identification of YSOs will depend upon forthcoming VLBA parallax measurements.

\section{Discussion} \label{sec:dis}
\subsection{Radio properties of the YSO population}
\begin{figure}[c]
  \centering
    \includegraphics{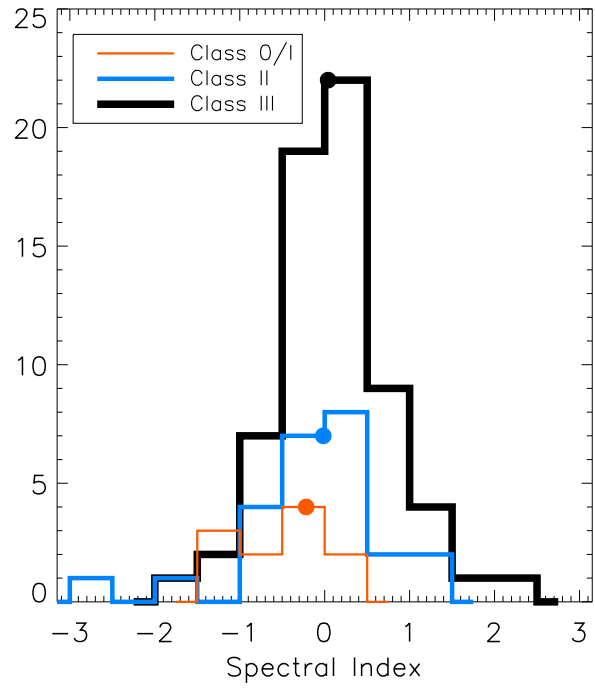}
      \caption{Spectral index distribution of the known and classified YSOs in our sample. Typical uncertainties of the spectral index 4.5 -- 7.5 GHz are 0.6. Circles show the median value for each SED class.\label{fig:spn}}
\end{figure}

We analyzed the radio properties of objects that have been previously identified as YSOs with a known Spectral Energy Distribution (SED) class. We compared the spectral indices in Figure \ref{fig:spn}. Unlike \cite{Dzib2013} who found the spectral index to be more negative for more evolved sources, we found no statistically significant difference in the spectral index between stars of different classes, and the median value for all three evolutionary classifications is consistent with zero.

\begin{figure}[c]
  \centering
    \includegraphics{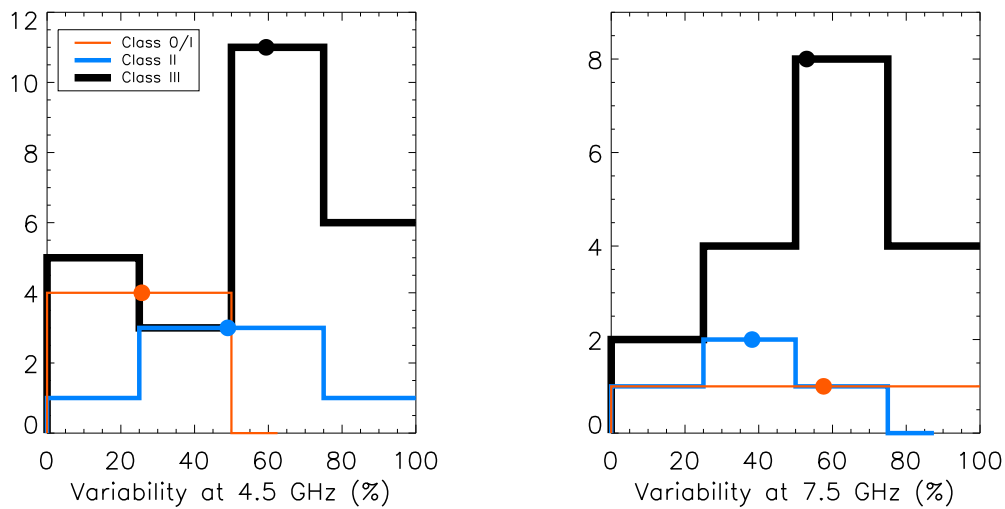}
      \caption{4.5 and 7.5 GHz variability of the known and classified YSOs in our sample. Only the sources which were detected at all three epoch are includeds. Typical uncertainties in variability are 25\%. Circles show the median value for each SED class.\label{fig:var}}
\end{figure}

Figure \ref{fig:var} shows variability of the classified YSOs. This is not representative of the total variability, as we do not include upper limits on the sources that were not detected in one or more epochs, which would skew distribution to be more variable for all of the classifications. Rather, we look at the relative distribution between different SED classes. As in \cite{Dzib2013}, we found that more evolved YSOs tend to be on average more variable as compared to their younger counterparts. This is likely due to non-thermal gyrosynchrotron emission becoming more prevalent over thermal bremsstrahlung radiation for older YSOs \citep{feigelson1999}.

\subsection{The radio - X-ray luminosity relation}

\begin{figure}[c]
  \centering
    \includegraphics{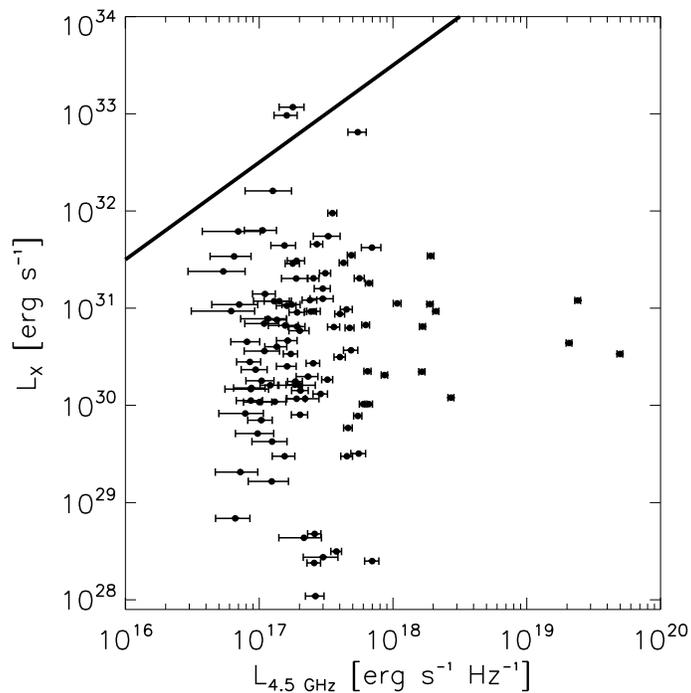}
      \caption{X-ray luminosity as a function of radio luminosity for stars in our sample. The black line is the approximate G\"{u}del-Benz relation. \label{fig:xray}}
\end{figure}

\cite{gb1993} found an empirical relation between X-ray and radio luminosity for magnetically active stars, suggesting that mechanisms that drive emission in both of these wavelengths are related. We compare our results to the G\"{u}del-Benz relation.

\[\dfrac{L_X}{L_R}\approx10^{15.5\pm1.5}\]

\noindent in Figure \ref{fig:xray}. Of 148 known YSOs in our sample, 114 have cited X-ray luminosities in either 3XMM-DR4 Source Catalogue \citep{3xmm} or in COUP Survey \citep{coup}. We did not include other X-ray surveys for a more consistent sample. We corrected all luminosities to the adopted distance of 414 pc \citep{menten2007}.  As \cite{Dzib2013} found in their analysis of the Ophiuchus region and \cite{forbrich2013} found in their studies of the ONC, we find that the X-ray emission of YSOs in our sample is underluminous compared to the G\"{u}del-Benz relation.

\section{Comments on some individual sources}\label{sec:sour}
\subsection{GMR F}
GMR F = GBS-VLA J053518.37$-$052237.4 is located in the central ONC region, not far from the Trapezium cluster. It has been identified in several multi-wavelength surveys and is a known variable star, with infrared photometry consistent with being a Class III YSO \citep{Megeath2012}. It emits non-thermally at radio wavelengths, and it was one of the four stars used by \cite{menten2007} to measure distance to the Trapezium with VLBA.

\begin{figure}[c]
  \centering
    \includegraphics{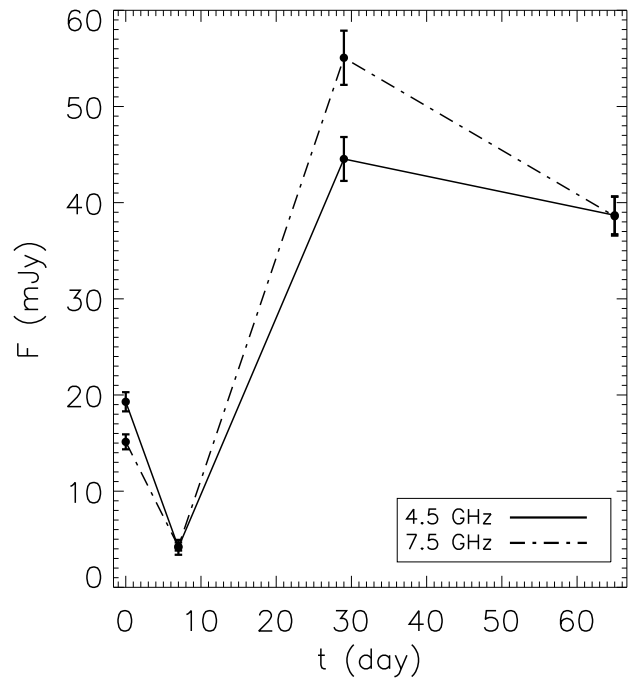}
      \caption{Light curve of GMR F at 4.5 and 7.5 GHz from Julian Date 2455737 to 2455802.\label{fig:gmrf}}
\end{figure}

GMR F is one of the few sources which we detected in more than three epochs, as fields which contained it were split between two different maps. We observed it undergoing a strong flare (Figure \ref{fig:gmrf}), in which the flux has increased by the factor of 90\% at both 4.5 and 7.5 GHz. It displayed noticeable circular polarization that was particularly pronounced in the last epoch, after the flux of the source started to decline.

\subsection{GBS-VLA J054121.69$-$021108.3}
GBS-VLA J054121.69$-$021108.3 is a star located near NGC 2023. It has been previously observed by \cite{reipurth2004} at radio wavelengths using the VLA at 8.3 GHz. They measured its flux to be on order of 50 mJy; since it was far from the beam center they could not give more exact values. It was later identified by \cite{mookerjea2009} and \cite{Megeath2012} as a Class II YSO using data from \textit{Spitzer Space Telescope}. It reportedly has no optical or submillimeter counterparts.

\begin{figure}[c]
  \centering
    \includegraphics{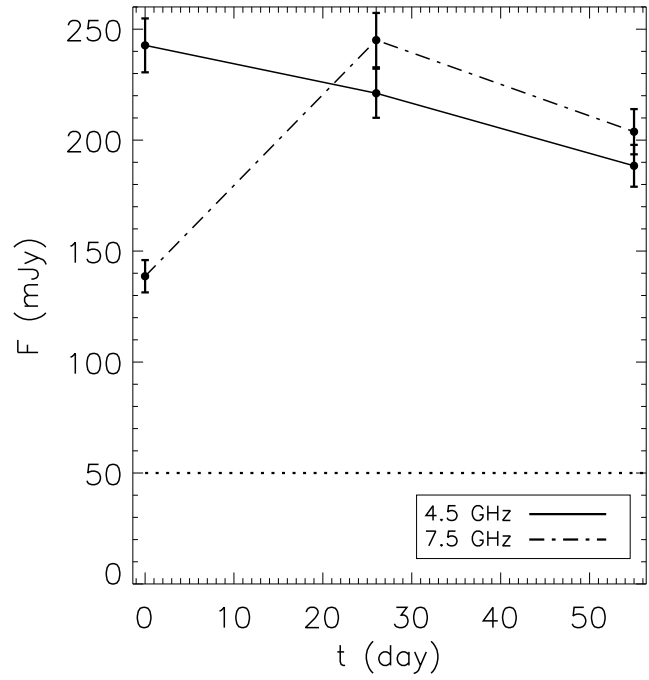}
      \caption{Light curve of GBS-VLA J054121.69$-$021108.3 at 4.5 GHz from Julian Date 2455746 to 2455801. Dotted line at 50 mJy shows the 8.3 GHz flux measured by \cite{reipurth2004} 3400 days prior to our observations.\label{fig:2023}}
\end{figure}

We observe the maximum flux of this source to be 243 and 245 mJy at 4.5 and 7.5 GHz, respectively (Figure \ref{fig:2023}). The maxima at each wavelengths did not occur at the same epoch; the maximum flux at 4.5 GHz was observed at the same time as the smallest flux at 7.5 GHz at just 137$\pm$7 mJy. This object does not show sufficient levels of variability ($>$50\%) over the short observation period for us to identify it as a variable star; however, combined with the detection by \cite{reipurth2004}, it undeniably shows longer term variability. This star may be a flaring source which we caught near its high state. Additionally, the centimeter flux density of this source is the highest ever reported in association with a young star.

This source was detected in the NRAO VLA Sky Survey (NVSS) at 1.4 GHz \citep{nvss} with a flux density of 98.4$\pm$3.9 mJy in observations made on 1995 March 5. We have analyzed the VLA archive observations of project AV187, finding a 1.4 GHz flux density of 72.6$\pm$2.3 mJy for observations taken on 1991 April 14 and 15. This result confirms the long-term variability of GBS-VLA J054121.69-021108.3.

\section{Conclusions} \label{sec:concl}

We report on radio observations in several regions in the Orion Molecular Cloud complex, namely $\lambda$ Ori, Lynds 1622, NGC 2068, NGC 2071, NGC 2023, NGC 2024, $\sigma$ Ori, ONC, and Lynds 1641. Our observations provide high sensitivity and angular resolution over 2.26 square degrees. We detected a total of 374 sources, of which 148 had been previously identified as YSOs. 86 unclassified sources exhibit radio properties that are consistent with those of young stars. These sources will be used as targets for future VLBA observations to determine the distance to and kinematics of these regions. These results will complement upcoming measurements from \textit{Gaia} and provide more concrete distances toward regions of high nebulosity or sources without optical counterparts.

\acknowledgments
This work was supported in part by the University of Michigan. L.L. is grateful to the von Humboldt Stiftung for financial support. L.L., S.D., L.F.R., G.N.O., G.P., and J.L.R. acknowledge the financial support of DGAPA, UNAM, and CONACyT, M\'{e}xico.  AIPS is produced and maintained by the National Radio Astronomy Observatory, a facility of the National Science Foundation operated under cooperative agreement by Associated Universities, Inc. This research has made use of the the SIMBAD database and VizieR catalogue access tool, operated at CDS, Strasbourg, France.

\input{t1.tex}
\input{t2.tex}
\input{t3.tex}
\input{t4.tex}
\input{t5.tex}



\end{document}

%% file: t1.tex
\begin{deluxetable}{ccccc}
\tabletypesize{\scriptsize}
\tablewidth{0pt}
\tablecaption{Dates of the observations of each map at each epoch as well as the observed regions. The total number of fields covering that region is given in parentheses.\label{tab:Dates}}
\tablehead{
\colhead{Map \#} & \colhead{Field} & \colhead{Epoch 1} & \colhead{Epoch 2} & \colhead{Epoch 3}\\
}
\startdata
Map 1 & $\lambda$ Ori (12), L1622 (3), NGC 2068 (15) & 07/04 & 07/30 & 08/25 \\
Map 2 & NGC 2068 (12),NGC 2023 (14), $\sigma$ Ori (4) & 07/04 & 07/30 & 08/28 \\
Map 3 & $\sigma$ Ori (7), ONC (23) & 07/04 & 07/30 & 08/28 \\
Map 4 & ONC (30) & 06/25 & 07/24 & 08/29 \\
Map 5 & ONC (30) & 07/02 & 07/24 & 08/29 \\
Map 6 & ONC (26), L1641-N (4) & 06/27 & 07/23 & 08/26 \\
Map 7 & L1641-N (12), L1642-C (8), L1641-S (10) & 07/03 & 07/23 & 08/27 \\
\enddata
\end{deluxetable}

%% file: t2.tex
\begin{deluxetable}{ccccccc}
\tabletypesize{\tiny}
\tablewidth{0pt}
\tablecaption{Radio sources detected in Orion\label{tab:all}}
\tablehead{
\colhead{GBS-VLA Name} & \colhead{$f_{4.5}$\tablenotemark{1} (mJy)} & \colhead{Var.\tablenotemark{2} (\%)} & \colhead{$f_{7.5}$\tablenotemark{1} (mJy)} & \colhead{Var.\tablenotemark{2} (\%)} & \colhead{Spectral Index\tablenotemark{3}} & \colhead{Variable?\tablenotemark{4}}\\
}
\startdata

J053450.56+100433.0&117.42$\pm$0.31$\pm$5.87&23.2$\pm$14.9&39.20$\pm$0.11$\pm$1.96&46.9$\pm$35.5&-2.27$\pm$0.15&n\\
J053451.13+100427.4&101.15$\pm$0.27$\pm$5.06&19.1$\pm$15.0&39.39$\pm$0.11$\pm$1.97&41.9$\pm$35.4&-1.95$\pm$0.15&n\\
J053428.42+100422.6&9.23$\pm$0.17$\pm$0.46&7.3$\pm$18.0&2.89$\pm$0.09$\pm$0.14&16.4$\pm$43.4&-2.41$\pm$0.16&n\\
J053512.43+100255.5&2.34$\pm$0.15$\pm$0.12&14.3$\pm$16.5&1.35$\pm$0.10$\pm$0.07&33.7$\pm$30.2&-1.14$\pm$0.25&n\\
J053434.92+095805.6&0.66$\pm$0.12$\pm$0.03&24.2$\pm$33.1&0.43$\pm$0.08$\pm$0.02&30.8$\pm$44.7&-0.90$\pm$0.56&n\\
J053455.56+095713.2&1.96$\pm$0.13$\pm$0.10&7.7$\pm$17.5&1.73$\pm$0.10$\pm$0.09&16.0$\pm$26.9&-0.25$\pm$0.24&n\\
J053503.54+095640.8&0.91$\pm$0.11$\pm$0.05&9.2$\pm$24.2&0.66$\pm$0.08$\pm$0.03&13.2$\pm$34.7&-0.66$\pm$0.38&n\\
J053451.35+095630.4&0.35$\pm$0.12$\pm$0.02&$>$12.3$\pm$57.5&0.29$\pm$0.08$\pm$0.01&---&-0.40$\pm$0.92&y\\
\enddata
\tablenotetext{1}{Highest detected flux between three epochs. The first quoted errors correspond to the statistical error calculated by \texttt{JMFIT}, and the second error is the 5\% absolute flux uncertainty.}
\tablenotetext{2}{The errors include the statistical and absolute uncertainty. $>$ indicates lack of detection at one of the epochs, and variability is calculated with flux of the remaining two epochs. --- with quoted flux indicates only a single epoch detection.}
\tablenotetext{3}{The errors include the statistical and absolute uncertainty.}
\tablenotetext{4}{Variability of at least 50\% or lack of detection at one or more epoch with the source located in the inner half of the beam.}
\end{deluxetable}

%% file: t3.tex
\begin{deluxetable}{ccc}
\tabletypesize{\scriptsize}
\tablewidth{0pt}
\tablecaption{Maximum observed plolarization for the sources at any epoch.\label{tab:pol}}
\tablehead{
\colhead{GBS-VLA Name} & \colhead{4.5 GHz polz (\%)} & \colhead{7.5 GHz polz (\%)}\\
}
\startdata
J053450.56+100433.0 & 6.8 (L) & 19.0 (L)\\
J053451.13+100427.4 & 6.9 (L) & 20.8 (L)\\
J054601.35+001422.9 & 7.0 (L) & 15.9 (L)\\
J054601.41+001422.8 & 7.6 (L) & ---\\
J054121.69$-$021108.3 & 2.7 (R) & 6.0 (R)\\
J053513.73$-$045201.9 & 2.6 (L) & 3.6 (L)\\
J053558.88$-$045537.7 & 2.8 (L) & 5.1 (L)\\
J053518.37$-$052237.4 & 2.5 (R) & 6.1 (R)\\
J053515.83$-$052314.1 & 3.4 (R) & 6.4 (R)\\
J053424.63$-$052838.5 & 1.4 (L) & 1.6 (L)\\
J053554.46$-$052437.1 & 10.5 (L) & 22.6 (L)\\
J053657.32$-$064802.1 & 12.5 (L) & ---\\
J053835.54$-$065913.7 & 5.3 (L) & ---
\enddata
\end{deluxetable}

%% file: t4.tex
\begin{deluxetable}{cccccc}
\tabletypesize{\tiny}
\tablewidth{0pt}
\tablecaption{Radio sources with known counterparts\label{tab:known}}
\tablehead{
\colhead{GBS-VLA Name} & \colhead{Object Type\tablenotemark{1}} & \colhead{X-ray\tablenotemark{2}} & \colhead{Optical\tablenotemark{3}}& \colhead{IR\tablenotemark{4}} & \colhead{Radio\tablenotemark{5}}\\
}
\startdata

J053428.42+100422.6&---&---&---&---&NVSS\\
J053530.67+095611.3&---&LOX 130&---&---&---\\
J053519.05+095454.9&YSO&LOX 98&BSM 9&HMC 4155&---\\
J053516.49+095437.3&Galaxy&---&---&2M&---\\
J053459.44+095312.0&---&LOX 51&---&---&---\\
J055403.03+014022.1&YSO (III)&---&TYC 117 1216 1&MGM&---\\
J054601.40+003206.9&---&---&---&MGM&---\\
J054630.76+003100.8&---&---&---&MGM&---\\
\enddata
\tablenotetext{1}{In parentheses listed YSO class, if known.}
\tablenotetext{2}{LOX=\cite{Franciosini2011}; 3XMM=\cite{3xmm}; SSM=\cite{skinner2009}; SGB=\cite{skinner2003}; CCS=\cite{Caballero2010}; RRS=\cite{ramirez2004}; TKT=\cite{tsujimoto2002}; COUP=\cite{coup}; SOXS=\cite{pillib2013}}
\tablenotetext{3}{BSM=\cite{Bayo2011}; TYC=\cite{tycho}; UCAC4=\cite{ucac4}; H97=\cite{Hill1997}; DRS=\cite{dario2009}; HHA=\cite{tina}}
\tablenotetext{4}{HMC=\cite{hernandez2010}; 2M=\citep{2mass}; MGM=\cite{Megeath2012}; PBZ=\citep{ramirez2012}; HHM=\cite{Hernandez2007}; MAD=\citep{Bouy2009}; ISOY=\cite{ysovar}; TKK=\cite{tsujimot2003}}
\tablenotetext{5}{NVSS=\cite{nvss}; G99=\cite{gibb1999}; RRA = \cite{reipurth2004}; RR=\cite{rodriguez1994}; RRC=\citep{reipurth1999}; GMR=\cite{gmr}; AVE=\cite{anglada1998}; ARC=\cite{avila2001}}
\end{deluxetable}

%% file: t5.tex
\begin{deluxetable}{cccc}
\tabletypesize{\tiny}
\tablewidth{0pt}
\tablecaption{Young stellar object candidates based just in their radio properties\label{tab:new}}
\tablehead{
\colhead{GBS-VLA Name} & \colhead{Var.$_{4.5}$ (\%)} & \colhead{Var.$_{7.5}$ (\%)} & \colhead{Spectral Index}\\
}
\startdata

J053450.56+100433.0&23.2$\pm$14.9&46.9$\pm$35.5&-2.27$\pm$0.15\\
J053451.13+100427.4&19.1$\pm$15.0&41.9$\pm$35.4&-1.95$\pm$0.15\\
J053451.35+095630.4&$>$12.3$\pm$57.5&---&-0.40$\pm$0.92\\
J053516.49+095437.3&43.3$\pm$65.2&---&-0.69$\pm$1.39\\
J053433.68+095308.8&$>$22.6$\pm$69.5&$>$26.3$\pm$78.9&-0.58$\pm$1.39\\
J053502.38+094911.6&53.1$\pm$18.3&$>$80.3$\pm$25.2&-1.10$\pm$0.31\\
J053557.21+094904.1&1.8$\pm$30.6&54.5$\pm$35.6&-1.09$\pm$0.45\\
J054731.55+003445.9&35.8$\pm$29.2&52.0$\pm$38.4&-0.07$\pm$0.46\\
\enddata
\end{deluxetable}